\documentclass[aps,prl,longbibliography,superscriptaddress,reprint]{revtex4-2}

\usepackage{graphicx}
\usepackage{textcomp}
\usepackage{amsmath}
\usepackage{multirow} 
\usepackage{physics} 
\usepackage{braket}
\usepackage{subfigure}   
\usepackage{floatrow}
\usepackage{verbatim}
\usepackage{color}
\usepackage{xcolor}
\usepackage[colorlinks,urlcolor=blue,citecolor=blue,linkcolor=blue]{hyperref}
\usepackage{soul}
\usepackage{tabularray}
\begin{document}

\title{Experimental Demonstration of the Timelike Unruh Effect with a Trapped-Ion System}

\author{Zhenghao Luo}
\affiliation{Laboratory of Spin Magnetic Resonance, School of Physical Sciences, Anhui Province Key Laboratory of Scientific Instrument Development and Application, University of Science and Technology of China, Hefei 230026, China}

\author{Yi Li}
\affiliation{Laboratory of Spin Magnetic Resonance, School of Physical Sciences, Anhui Province Key Laboratory of Scientific Instrument Development and Application, University of Science and Technology of China, Hefei 230026, China}
\affiliation{Hefei National Laboratory, University of Science and Technology of China, Hefei 230088, China}
\affiliation{National Advanced Talent Cultivation Center for Physics, University of Science and Technology of China, Hefei 230026, China}

\author{Xingyu Zhao}
\affiliation{Laboratory of Spin Magnetic Resonance, School of Physical Sciences, Anhui Province Key Laboratory of Scientific Instrument Development and Application, University of Science and Technology of China, Hefei 230026, China}
\affiliation{Hefei National Laboratory, University of Science and Technology of China, Hefei 230088, China}

\author{Zihan Xie}
\affiliation{Laboratory of Spin Magnetic Resonance, School of Physical Sciences, Anhui Province Key Laboratory of Scientific Instrument Development and Application, University of Science and Technology of China, Hefei 230026, China}
\affiliation{Hefei National Laboratory, University of Science and Technology of China, Hefei 230088, China}

\author{Zehua Tian}\email{tzh@hznu.edu.cn}
\affiliation{School of Physics, Hangzhou Normal University, Hangzhou 311121, China}
\affiliation{Laboratory of Spin Magnetic Resonance, School of Physical Sciences, Anhui Province Key Laboratory of Scientific Instrument Development and Application, University of Science and Technology of China, Hefei 230026, China}

\author{Yiheng Lin} \email{yiheng@ustc.edu.cn}
\affiliation{Laboratory of Spin Magnetic Resonance, School of Physical Sciences, Anhui Province Key Laboratory of Scientific Instrument Development and Application, University of Science and Technology of China, Hefei 230026, China}
\affiliation{Hefei National Research Center for Physical Sciences at the Microscale, University of Science and Technology of China, Hefei 230026, China}
\affiliation{Hefei National Laboratory, University of Science and Technology of China, Hefei 230088, China}

\date{\today}

\begin{abstract}
The Unruh effect predicts that an accelerated observer perceives the Minkowski vacuum as a thermal bath, but its direct observation requires extreme accelerations beyond current experimental reach. Foundational theory [Olson \& Ralph, Phys. Rev. Lett. 106, 110404 (2011)] shows that an equivalent thermal response, known as the timelike Unruh effect, can occur for detectors following specific timelike trajectories without acceleration, enabling laboratory tests with stationary yet time-dependent detectors. Here, we report a proof-of-principle demonstration of the timelike Unruh effect in a quantum system of trapped ion, where a two-level spin serves as the detector and is temporally coupled to the ambient field encoded in the ion's vibrational motion. Specifically, we study both excitation and emission dynamics of the detector moving along a spacetime trajectory in the future/past light cone, and demonstrate the thermal response of the detector to the Minkowski vacuum that resembles the Unruh effect. This work establishes a controllable tabletop platform for exploring relativistic quantum physics under accessible laboratory conditions.
\end{abstract}

\maketitle  

\begin{figure*}[t!]
	\begin{center}
	\includegraphics[width=1\columnwidth]{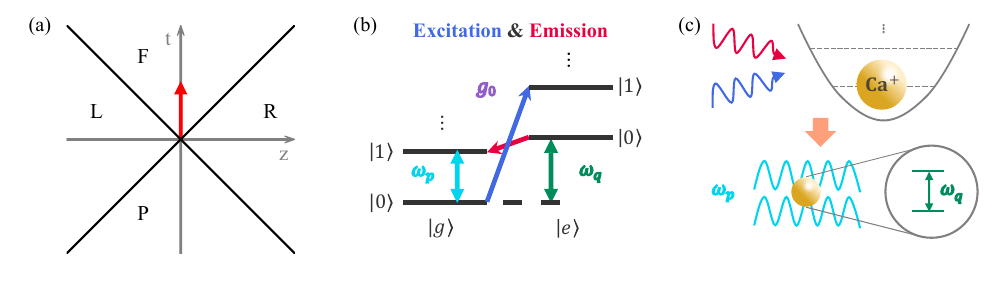}
	\caption{
        (a) A spacetime diagram separated into four quadrants: the left and right Rindler wedges (L and R), and the future and past light cones (F and P).
        The red arrow represents the spacetime trajectory of the detector.
        (b) Sketch of experimental model. It represents the model at the beginning $t=\tau=0$. Spin states are labeled $\{\ket{g},\ket{e}\}$, while photon numbers are labeled $\ket{n=0, 1, …}$ and plotted vertically. The excitation and emission process corresponds to blue and red arrows respectively. The spin frequency $\omega_q$ is bigger than the field frequency $\omega_p$ and the coupling strength is $g_0$.
        (c) A trapped $^{40}{\rm Ca}^{+}$ ion in the ground state of vibrational level applied with two customized red and blue sideband lasers, could be equalized as a two-level detector, with internal transition frequency $\omega_q$, coupling with a single mode photon field of frequency $\omega_p$ in the vacuum.
	}
	\label{fig:Model}
	\end{center}
\end{figure*}

The Unruh effect~\cite{unruh_notes_1976} is one of the fascinating phenomena predicted by the application of the theory of relativity to quantum mechanics, and plays a key role in the exploration of the not-yet-understood physics of quantum gravity.
It arises out of the basic property of the quantum vacuum, wherein the vacuum state of the field can be expressed as an entangled state between two sets of modes, which respectively span the left and right (L and R) Rindler wedges in the Minkowski ($\mathcal{M}$) spacetime~\cite{crispino_unruh_2008, birrell_quantum_1982} (see Fig.~\ref{fig:Model}(a)).
Since a uniformly accelerated observer can only measure one set of Rindler modes, tracing out the unobserved modes leads to the prediction that such an observer will see a thermalized vacuum.
The Unruh temperature sensed by an accelerating observer is exceedingly small; for example, achieving a temperature of 1~K would require an acceleration on the order of $10^{20}~\mathrm{m/s^2}$.
Although various experimental settings have been proposed to detect the Unruh effect~\cite{bell_electrons_1983, yablonovitch_accelerating_1989, chen_testing_1999, schutzhold_signatures_2006, schutzhold_tabletop_2008, martin-martinez_using_2011, guedes_spectra_2019, lochan_detecting_2020, gooding_interferometric_2020, arya_geometric_2022, tian_probing_2022,stargen_cavity_2022, arya_lamb_2023}, the intractable, but required, relativistic motion is still the main obstacle to verifying the Unruh effect in practice and the relevant researches only stay at the level of quantum simulation~\cite{BEC,NMR}.

Foundational theory~\cite{olson_entanglement_2011,olson_extraction_2012} has proposed the timelike Unruh effect: an Unruh-DeWitt detector confined within the future or past (F or P) light cone exhibits the same thermal response to the quantum vacuum as an accelerated detector in a Rindler wedge.
This occurs due to an identical entanglement structure between quantum fields in the light-cone regions, which is termed timelike entanglement, compared with that in the Rindler wedges (see Refs.~\cite{olson_entanglement_2011, higuchi_entanglement_2017, ueda_entanglement_2021} and Fig.~\ref{fig:Model}(a)).
A detector following a trajectory within one light cone interacts with the field and, after tracing out unobserved modes, perceives the Minkowski vacuum as thermal.
Proposals~\cite{quach_berry_2022, tian_using_2023} suggest observing this effect without spatial acceleration by continuously modulating the detector’s energy gap~\cite{olson_entanglement_2011}.
This approach enables experimental study of Unruh physics, as well as quantum field theory in curved spacetime~\cite{crispino_unruh_2008,
birrell_quantum_1982, olson_entanglement_2011, olson_extraction_2012}.
However, its experimental realization, or even the proof-of-principle demonstration has remained elusive.

Here, we report a proof-of-principle demonstration of the timelike Unruh effect using a single trapped ion.
The trapped-ion system~\cite{leibfried_quantum_2003} is a well-established quantum platform that has been widely proposed for studying relativistic quantum effects, including Gibbons–Hawking radiation~\cite{alsing-ion-2005, menicucci-single-2007, menicucci-simulating-2010}, cosmological particle creation~\cite{schutzhold-analogue-2007, wittemer-phonon-2019}, and Hawking radiation~\cite{horstmann-hawking-2011, horstmann-hawking-2010, tian_testing_2022}.
In our experiment, the spin–motion interaction of ion corresponds to that between a detector and a single-mode quantum field.
By engineering the spin–motion interaction via laser control, we observe the thermal response to vacuum experienced by an inertial detector coupled only to a quantum field in the future light cone.
Through excitation and emission processes, we analyze the detector’s dynamics and demonstrate its thermalization with the field vacuum.
Our results show clear agreement with theoretical predictions of timelike Unruh radiation and provide insights for future experimental observations of relativistic quantum phenomena.

\textit{Theoretical model.}---%
The $\mathcal{M}$ spacetime is separated into four quadrants F, P, L, R, (shown in Fig.~\ref{fig:Model}(a)), and an accelerated observer is confined in one of these regions. Here we focus on the future region scenario, and correspondingly 
the future-past (FP) coordinates in the future light cone $(\tau,\zeta)_\text{FP}$ transform the usual $\mathcal{M}$ coordinate $(t,z)_\text{M}$ as $t = \alpha^{-1}e^{\alpha\tau}\cosh{(\alpha\zeta/c)}, z = c\alpha^{-1}e^{\alpha\tau}\sinh{(\alpha\zeta/c)}$~\cite{olson_entanglement_2011}, where $\alpha=a/c$ is corresponding effective acceleration in units of $\text{s}^{-1}$, while $a$ is real acceleration and $c$ is the speed of light. The thermal response of the timelike Unruh effect does not require direct observation of the correlation of the modes of the future and past light cones~\cite{olson_entanglement_2011}. As such, we will restrict ourselves to a detector in the future light cone with the world line $(\tau,0)_\text{FP}$, then the transformation can be simplified as
\begin{equation}
    t = \alpha^{-1}e^{\alpha\tau},~~z =\zeta = 0.
    \label{Eq:Transformation_t}
\end{equation}

For a detector moving along the above worldline in Eq.~\eqref{Eq:Transformation_t}, the corresponding Schr\"odinger equation in its intrinsic framework (FP coordinates) takes the following form~\cite{olson_entanglement_2011}:
\begin{eqnarray}\label{S-equation-tau}
i\hbar\frac{\partial\psi}{\partial\tau} = \hat{H}^\text{FP}\psi
= (\hat{H}_d+e^{\alpha\tau}\hat{H}_I)\psi,
\end{eqnarray}
where $\hat{H}_d = \hbar\omega_q\hat{\sigma}_z/2$ denotes a two-level detector with constant energy gap in its intrinsic framework, and
$\hat{H}_I = \hbar\omega_p\hat{b}^\dagger\hat{b} + \hbar g_0\chi(\tau)\hat{\sigma}_x(\hat{b}+\hat{b}^\dagger)$ is the standard interaction term for an Unruh-DeWitt detector, characterizing that the detector is coupled to a single-mode photon field.
Here, $\omega_q,\omega_p,g_0$ are respectively the energy spacing of the detector, the photon frequency and the coupling strength, while $\hat{\sigma}_i$ is the Pauli operator and $\hat{b}~(\hat{b}^\dagger)$ represents the annihilation (creation) operator of the photon field mode.
Additionally, $\chi(\tau)\in [0,1]$ in the coupling term denotes the switching function, which signifies the duration of interaction between the detector and the field~\cite{junker-adiabatic-2002, louko-transition-2008}.
Note that this kind of switching function has been commonly taken into account when exploring the quantum field in curved space through Unruh-DeWitt detector, for example in ~\cite{Alejandro-harvesting-2015,Eduardo-causality-2015,BLHu-relativistic-2012,Alejandro-Then-2007,louko-how-2006}  and the references therein.
Therefore, the total Hamiltonian of the model in the FP coordinates reads
\begin{equation}
    \begin{aligned}
        \hat{H}^\text{FP}(\tau)
        =\frac{\hbar\omega_q}{2}\hat{\sigma}_z
        +\hbar\omega_p e^{\alpha\tau}\hat{b}^\dagger\hat{b}
        +\hbar g_0\chi(\tau)e^{\alpha\tau}\hat{\sigma}_x(\hat{b}+\hat{b}^\dagger).
    \end{aligned}
\end{equation}
Thus, in the accelerated observer's frame, the field frequency and the coupling involve time-dependent Doppler shifts, i.e., $\omega^\prime_p = \omega_p e^{\alpha\tau}$ and $g^\prime_p = g_0 e^{\alpha\tau}$, while the detector's energy spacing is fixed. In the laboratory framework ($\mathcal{M}$ coordinates), the above Schr\"odinger equation Eq.~\eqref{S-equation-tau} corresponds to 
\begin{eqnarray}\label{S-equation-t}
i\hbar\frac{\partial\psi}{\partial t} = i\hbar\frac{\partial\psi}{\partial\tau} \cdot \frac{\partial\tau}{\partial t}=\frac{\hat{H}^\text{FP}}{\alpha t} \psi=\hat{H}^\text{M}\psi.
\end{eqnarray}
Therefore, the Hamiltonian $\hat{H}^\text{M}$ in the $\mathcal{M}$ coordinates reads
\begin{equation}
\begin{aligned}
    \hat{H}^\text{M}(t)
    &= \cfrac{1}{\alpha t} \cfrac{\hbar\omega_q}{2}\hat{\sigma}_z + \hbar\omega_p \hat{b}^\dagger \hat{b}
    + \hbar g_0 \chi(t) \hat{\sigma}_x(\hat{b}+\hat{b}^\dagger).
    \label{Eq:H_M_theory}
\end{aligned}
\end{equation}
Thus, in the frame of laboratory observer, the detector's energy spacing has a temporal scaling, $\omega^\prime_q=\omega_q/\alpha t$, while the field frequency and the coupling strength are invariant. Seen from Eq.~\eqref{Eq:H_M_theory}, in order to realize a accelerated Unruh-DeWitt detector in the future light cone in the laboratory framework, large energy-level spacing of the detector at small $t$ is needed, which seems to be 
not physically realizable. Therefore, we apply a time translation $t\to t+C/\alpha$.
The Hamiltonian we construct for experiment reads
\begin{equation}
    \hat{H}^\text{M}_\text{exp}(t) = \frac{1}{\alpha t+C} \cfrac{\hbar\omega_q}{2}\hat{\sigma}_z + \hbar\omega_p \hat{b}^\dagger \hat{b}
    + \hbar g_0 e^{-t/T_d} \hat{\sigma}_x(\hat{b}+\hat{b}^\dagger).
    \label{Eq:H_M_exp}
\end{equation}
The scaling factor of spin frequency turns into $1/(\alpha t+C)$ and the initial ($t=0$) spin spacing is $\omega_q/C$, which is controllable in experiment.
Specially when $C=0$ it returns to Eq.~\eqref{Eq:H_M_theory} and when $C=1$ the initial spin spacing becomes $\omega_q$.

To illustrate the thermalization of the timelike Unruh effect, we consider both the spontaneous excitation and emission processes~\cite{Soda_Acceleration_2022} of an accelerated detector in the Minkowski field vacuum shown in Fig.~\ref{fig:Model}(b).
As such, we prepare the initial state of detector and field at $\ket{\psi_i}$, and observe the corresponding transition probabilities to the final state $\ket{\psi_f}$.
According to time-dependent perturbation theory, the transition probabilities to the first order can be evaluated as
\begin{equation}
    P_{fi}(t) = \abs{\int_0^t{\dd t \bra{\psi_f}\hat{V}^\text{M}(t)\ket{\psi_i}}}^2,
    \label{Eq:P_fi}
\end{equation}
where $\hat{V}^\text{M}(t) = \hat{U}^\dagger(t) \hat{H}^\text{M}_{exp}(t) \hat{U}(t)$ is the corresponding Hamiltonian in the interaction picture~\cite{SM}.
Specifically, $P_{fi}$ is denoted as $P_\text{exc}$ for the excitation process with $\ket{\psi_i}=\ket{g,0},\ket{\psi_f}=\ket{e,1}$ and as $P_\text{emi}$ for the emission process with $\ket{\psi_i}=\ket{e,0},\ket{\psi_f}=\ket{g,1}$.
We consider a coupling would damp exponentially and vanish in the end, i.e. $\chi(t)=\exp(-t/T_d)$, where $T_d$ is regarded as damping time.
Such a choice of damping would stabilize evolution gradually, and in an ideal case with $C=0$ and $\omega_p T_d \gg 1$, the final transition probability $P_\text{final} = P_{fi}(+\infty)$ for excitation and emission processes are respectively found to be~\cite{SM}
\begin{equation}
    \begin{aligned}
        P_\text{exc,final}
        &= \frac{g_0^2}{\omega_p^2}\frac{2\pi\omega_q}{\alpha}
            \times\frac{1}{\exp(\frac{2\pi\omega_q}{\alpha})-1}\\
        &= \frac{g_0^2}{\omega_p^2}\frac{\hbar\omega_q}{k_B T_U}
            \times\frac{1}{\exp(\frac{\hbar\omega_q}{k_B T_U})-1}\\
        &=\frac{g_0^2}{\omega_p^2}\times \frac{1/\beta}{\exp(1/\beta)-1},\\
        P_\text{emi,final}
        &=\frac{g_0^2}{\omega_p^2}\times \frac{1/\beta}{1-\exp(-1/\beta)},
        \label{Eq:P_final}
    \end{aligned}
\end{equation}
where $T_U=\hbar \alpha / 2\pi k_B$ is the Unruh temperature and $\beta=k_B T_U/\hbar\omega_q$ is the dimensionless Unruh temperature.
This corresponds to the Bose-Einstein (BE) distribution at the Unruh temperature during the process, indicating the thermalization of Unruh effect~\cite{ben-benjamin_unruh_2019}.

As a comparison, when we remove the acceleration term $1/(\alpha t+C)$ in Eq.~\eqref{Eq:H_M_exp}, it becomes the standard detuned Rabi model and we find almost no excitation or emission through numerical simulations.
We also verify such a time translation in our parameter of choice would not influence our results.
These points are detailed in~\cite{SM}.

Additionally, to explicitly display the relationship between the results and Unruh temperature, the effective transition probabilities could be defined as
\begin{equation}
    \begin{aligned}
        P_\text{eff}
        &=P_\text{final}\times\frac{\omega^2_p}{g^2_0}\\
        &= \frac{1}{\beta} \times
        \left\{  
             \begin{array}{lr}
             1/[\exp(1/\beta)-1] & (\text{excitation})\\
             1/[1-\exp(-1/\beta)]& (\text{emission})\\
             \end{array}  
        \right.,
    \end{aligned}
    \label{Eq:P_eff}
\end{equation}
which only depend on the parameter $\beta$.
Furthermore, their ratio yields
\begin{equation}
    \label{Eq:eta}
    \eta=P_\text{emi,eff}/P_\text{exc,eff}=\exp(1/\beta),
\end{equation}
which corresponds to Einstein thermal equilibrium conditions.
The above derivation is a simplified expression of Eq.~\eqref{Eq:P_final} and introduces the effective transition probabilities and their ratio, which depend uniquely on the parameter $\beta$.

In summary, the evolution of this detector-field system as in Eq.~\eqref{Eq:H_M_exp} will finally tend to thermal equilibrium due to the damping term.
The final transition probability indicates acceleration-dependent thermal distribution, which is a typical property of Unruh effect.
This guides us to demonstrate the timelike Unruh effect by studying the functional dependence of the transition probability on scanned acceleration values.

\textit{Experimental setup.}---%
The experiment employs a single $^{40}\text{Ca}^{+}$ ion confined in a linear Paul trap. 
The axial motional mode, quantized as a harmonic oscillator with frequency $\omega_z / 2\pi = 1.096$~MHz, 
generates phonon states $\{\ket{n} \mid n = 0,1,\ldots\}$ that are regarded as a bosonic quantum field.
We encode a qubit in the electronic states: $\ket{g} \equiv \ket{^{2}\text{S}_{1/2}, m_j = +1/2}$ and $\ket{e} \equiv \ket{^{2}\text{D}_{5/2}, m_j = +1/2}$.
The electric quadrupole transition between these states, driven by a 729-nm laser, corresponds to the detector frequency $\omega_0$. 
This realizes an Unruh-DeWitt detector with Hilbert space $\mathcal{H} = \text{span}\{\ket{g},\ket{e}\} \otimes \{\ket{n}\}$, characterizing a two-level system coupled to a bosonic field (shown in Fig.~\ref{fig:Model}(c)).
To satisfy first-order approximation constraints in Eqs.~\eqref{Eq:P_fi} and \eqref{Eq:P_final}, our experimental parameters must yield small yet measurable values of $P_\text{final}$.
We therefore fix: $\omega_q/2\pi = 200~\text{kHz},\quad g_0/2\pi = 5~\text{kHz},\quad C=1,\quad T_d = 0.2~\text{ms}$ while scanning acceleration $\alpha$ from $1\,\text{MHz}$ to $1\,\text{GHz}$.
Additionally, we set $\omega_p/2\pi = 25,\ 50~\text{kHz}$ for different implementations (detailed in~\cite{SM}).

At the beginning, we initialize the ion state to near $\ket{g,0}$ via Doppler cooling followed by sideband cooling.
For the emission experiment, we apply an additional $\pi$-pulse to prepare the initial state $\ket{e,0}$.
To implement the Hamiltonian in Eq.~\eqref{Eq:H_M_exp}, we simultaneously apply custom red and blue sideband lasers with identical exponentially damped intensities~\cite{SM}.
Their frequencies are tuned to $[\omega_0 - \omega_q \ln(\alpha t + 1)/(\alpha t)] \pm (\omega_z - \omega_p)$.
This is technically achieved using an arbitrary waveform generator driving an acousto-optic modulator.
During evolution, population primarily distributes between $\ket{g,0}$/$\ket{e,0}$ and $\ket{e,1}$/$\ket{g,1}$ for excitation/emission processes.
Consequently, the spin-state populations $P_g$/$P_e$ directly reflect the transition probability $P_{fi}$, allowing phonon-state detection to be omitted~\cite{SM}.
Final spin distributions are measured via spin-dependent fluorescence detection by a photomultiplier tube.

\begin{figure}[tbp]
	\begin{center}
	\includegraphics[width=1\columnwidth]{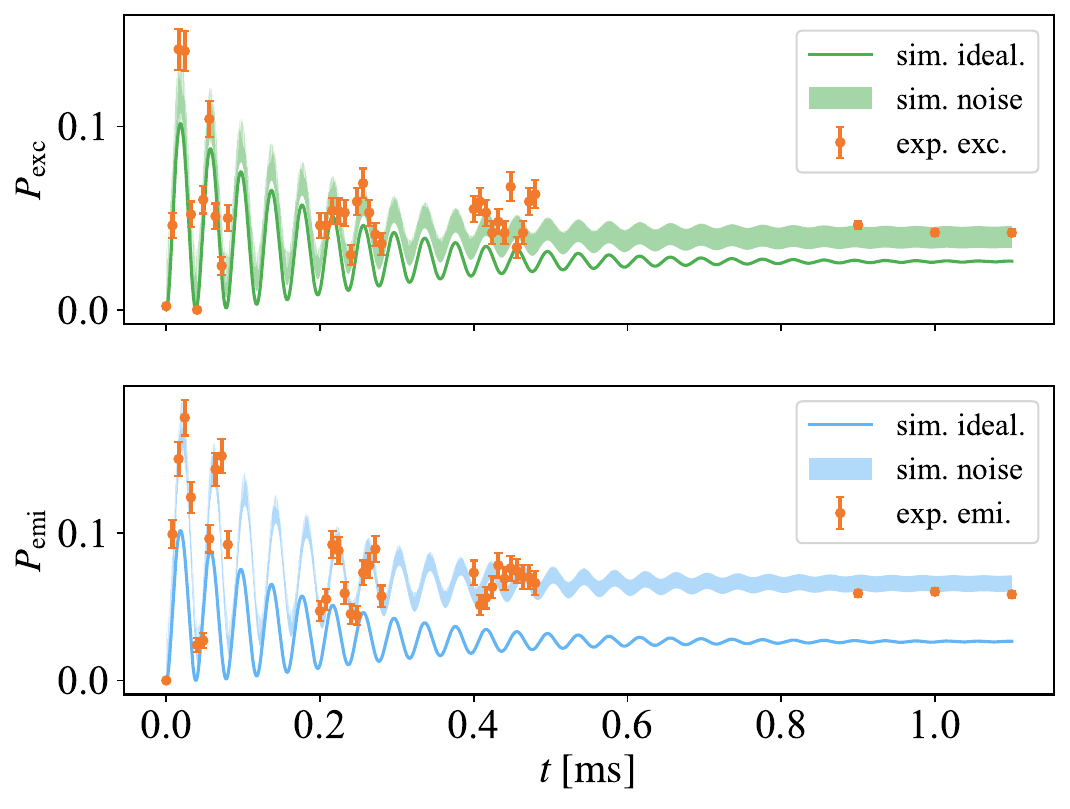}
	\caption{
            Evolution of transition probability during excitation (top) and emission (bottom) processes.
            Here, $\alpha = 10~\text{MHz}, \omega_p/(2\pi)=25~\text{kHz}$ and three time periods are set as 0-80, 200-280 and 400-480~\textmu s.
            The experimental points at $t>0.8~\text{ms}$ are detected with $10^4$ repetitions while others are detected with $10^3$ repetitions.
        }
	\label{fig:Evolution data}
	\end{center}
\end{figure}

\textit{Evolution of the detector-field system.}---%
We demonstrate the detector-field's evolution by preparing the initial states $\ket{g,0}$ and 
$\ket{e,0}$, and respectively measuring the corresponding excitation and emission probabilities, namely $P_\text{exc}$ and $P_\text{emi}$.
Fig.~\ref{fig:Evolution data} shows experimental and simulated results.
Comparative observation across three representative time intervals captures the oscillatory damping effect, which results from the switching function in the detector-field coupling.
Moreover, we perform measurements at about $t=5T_d=1~\text{ms}$ with more repetitions and observe stabilized transition probability, which we therefore determine as the final transition probability $P_\text{final}$.
Though experimental results deviate from ideal simulations (solid line), they align with simulations with experimental noise (error band).
Discrepancies arise primarily from excited thermal phonon state due to imperfect cooling.
Additionally, during repetitions initial phonon-state instability and random relative phase of two lasers lead to wide error band~\cite{SM}.
In summary, the full evolution exhibiting damped oscillations in transition probability reaches a steady state, consistent with the prediction that the detector equilibrates thermally via interaction with the vacuum field.

\begin{figure}[tbp]
	\begin{center}
	\includegraphics[width=1\columnwidth]{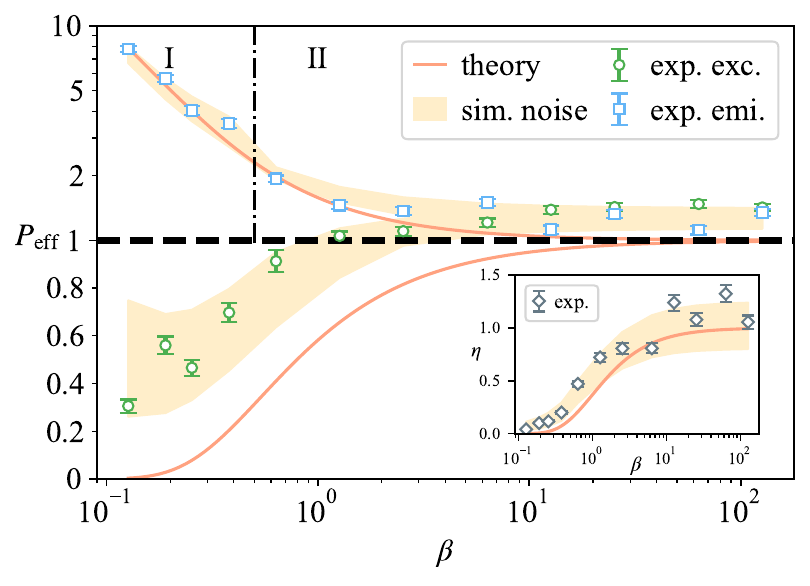}
	\caption{
        Effective transition probability $P_\text{eff}$ with various $\beta$.
        Experimental data for the excitation (emission) process are shown as green circles (blue squares), with corresponding error bars indicating one standard deviation. Theoretical predictions from Eq.~\eqref{Eq:P_eff} are plotted as orange solid lines, as separated by the horizontal dashed line $P_\text{eff}=1$.
        The emission part is implemented with $\omega_p/(2\pi)=50~\text{kHz}$ and $25~\text{kHz}$, labeled with zones I,II respectively.
        The simulation results considering effects from experimental noise are displayed by yellow bands.
        The inset shows the transition probability ratio $\eta$ for various values of $\beta$.
        }
	\label{fig:P_eff}
	\end{center}
\end{figure}

\textit{Demonstration of the timelike Unruh effect.}---%
After characterizing the system dynamics, we measure $P_\text{final}$ as a function of acceleration, which is expected to embody thermalization of the time-like Unruh effect.
The effective transition probability $P_\text{eff}$ in Eq.~\eqref{Eq:P_eff} is shown in Fig.~\ref{fig:P_eff}.
For the emission process, two photon frequencies $\omega_p$ are set (zones I, II) to satisfy the approximation $P_\text{emi,final} \ll 1$, and this change would not affect the results of $P_\text{eff}$~\cite{SM}.
Theory (red curves) predicts $P_\text{exc,eff} \leq 1 \leq P_\text{emi,eff}$, dividing the figure at $P_\text{eff}=1$; both curves converge with increasing acceleration.
Experiment deviates slightly from ideal theory but matches simulations with experimental noise (error bands, same as that in Fig.~\ref{fig:Evolution data}) and retains the predicted functional form.
In the inset, the Einstein thermal equilibrium condition $\eta \equiv P_\text{exc,eff}/P_\text{emi,eff}=\exp(-1/\beta)=\exp(-\hbar\omega_q/k_B T)$, with $T = T_U$ (Unruh temperature), is shown.  
These results collectively embody the acceleration-dependent Unruh thermalization, demonstrating agreement with theoretical predictions and providing conclusive experimental demonstration of the timelike Unruh effect.

\textit{Conclusion.}---%
In summary, we demonstrate the proof of principle of the timelike Unruh effect in a trapped-ion system.
We construct an Unruh-DeWitt detector coupled to a single-mode quantum field vacuum confined within the future light cone along the worldline $(\tau,0)_\text{FP}$.
Through observing detector-field dynamics and extract final steady states, we experimentally verify the thermal character of the effect, which is manifested through Bose-Einstein statistics and thermal equilibrium at Unruh temperature $T_U = \hbar\alpha/2\pi k_B$.
This work establishes a novel methodology for exploring quantum field theory in curved spacetime. Based on the equivalence principle, our proof-of-principle demonstration confirms the thermal nature of Hawking radiation in gravitational fields.

Generally, the phonon-for-photon substitution we utilize establishes trapped-ion system as a promising platform for studying a wider range of quantum field theory phenomena in controllable laboratory settings~\cite{alsing-ion-2005,horstmann-hawking-2010,horstmann-hawking-2011,pedernales_dirac_2018}.
Our work paves a way to demonstrations with multi-mode fields for structure spectrum. In addition, the utilization of phonon fields alleviates the need for fast control comparable to the speed of light, assisting the revelation of the dynamics.

\begin{acknowledgments}
\emph{Acknowledgments.}---%
We thank T. Ralph for useful discussion and helpful instruction. The USTC team acknowledges support from the National Natural Science Foundation of China (Grant No.~92165206), the Chinese Academy of Sciences (Grant No.~XDB1300000), and Innovation Program for Quantum Science and Technology (Grant No.~2021ZD0301603). Z.T was supported by the scientific research start-up funds of Hangzhou Normal University: 4245C50224204016.
\end{acknowledgments}

\onecolumngrid   
\vspace{5em}

\begin{center}
    \textbf{\fontsize{12}{12}\selectfont Supplementary Materials}
\end{center}

\twocolumngrid   

\appendix
\setcounter{figure}{0}
\renewcommand{\thefigure}{S\arabic{figure}}  
\setcounter{table}{0}
\renewcommand{\thetable}{S\arabic{table}}    
\setcounter{equation}{0}
\renewcommand{\theequation}{S\arabic{equation}} 

\renewcommand{\theHfigure}{S.\arabic{figure}} 
\renewcommand{\theHtable}{S.\arabic{table}}
\renewcommand{\theHequation}{S.\arabic{equation}}

\section{Spontaneous excitation and emission of the detector in Unruh effect}
The Hamiltonian of the whole system (detector plus the field) in the $\mathcal{M}$ coordinate reads ($\hbar=1$)
\begin{eqnarray}
\label{Eq:H_M_theory}
\hat{H}^\text{M}(t)=\frac{1}{\alpha t}\frac{1}{2}\omega_q\hat{\sigma}_z+\omega_p \hat{b}^\dagger\hat{b}+g_0 \chi(t)\hat{\sigma}_x(\hat{b}^\dagger+\hat{b}),
\end{eqnarray}
where $\chi(t)$ is the switching function~\cite{junker-adiabatic-2002, louko-transition-2008}: the interaction takes place only when $\chi$ nonvanishing, and because $\chi$ has compact support the interaction has finite duration.
In the interaction picture, with the rotation operator $\hat{U}(t) = \exp{-i\frac{\omega_q}{2\alpha}[\ln{(\alpha t)} - \alpha t]\hat{\sigma}_z} \exp(-i\omega_p t \hat{b}^\dagger\hat{b})$, the whole system is dominated by the Hamiltonian 
\begin{eqnarray}
\begin{aligned}
    \hat{V}^\text{M}(t)
    &= \hat{U}^\dagger(t) \hat{H}^\text{M}(t) \hat{U}(t)
    \\
    &=g_0\chi(t)(\hat{b}^{\dagger}e^{i\omega_pt}+\hat{b}e^{-i\omega_pt}) 
    \\
    &\quad\times[\hat{\sigma}_+ e^{i\omega_q\ln{(\alpha t)}/\alpha} + \hat{\sigma}_- e^{-i\omega_q\ln{(\alpha t)}/\alpha}].
\end{aligned}
\end{eqnarray}

According to time-dependent perturbation theory, the transition probabilities $P_{fi}$ to the first order can be evaluated as 
\begin{equation}
\label{Eq:P_fi}
    P_{fi}(t) = \abs{\int_0^t{\dd t \bra{\psi_f}\hat{V}^\text{M}(t)\ket{\psi_i}}}^2,
\end{equation}
where $\ket{\psi_i}$ is the initial state and $\ket{\psi_f}$ is the final state, and we note that this approximation requires $P_{fi}\ll1$.
For the excitation process, the initial state is $\ket{\psi_i}=\ket{g,0}$ and then $\bra{\psi_f}\hat{V}^\text{M}(t)\ket{\psi_i}\neq0$ only when $\ket{\psi_f}=\ket{e,1}$.
Thus the excitation probability is defined by $$P_\text{exc}(t) = \abs{\int_0^t{\dd t \bra{e,1}\hat{V}^\text{M}(t)\ket{g,0}}}^2.$$
Similarly, the emission probability is defined by
$$P_\text{emi}(t) = \abs{\int_0^t{\dd t \bra{g,1}\hat{V}^\text{M}(t)\ket{e,0}}}^2.$$

Here, we take the switching function as $\chi(t)=\exp(-t/T_d)$ and calculate the transition probability (including both the excitation and emission).
For excitation, it reads
\begin{eqnarray}
\begin{aligned}
    P_\text{exc,final}
    &=\abs{\int_0^\infty\,\dd t \bra{e,1}\hat{V}^\text{M}(t)\ket{g,0}}^2
    \\
    &=g_0^2\abs{\int^\infty_{0}\dd t e^{-t/T_d}e^{i\omega_pt} e^{i\omega_q\ln(\alpha t)/\alpha}}^2
    \\
    &=g_0^2\abs{\int^\infty_{0}\dd t e^{-(1/T_d-i\omega_p)t}(\alpha t)^{i\omega_q/\alpha}}^2.
\end{aligned}\label{Pe1}
\end{eqnarray}
From the definition of Gamma function,$$\frac{1}{\mu}\Gamma\bigg(\frac{\nu}{\mu}\bigg)\frac{1}{z^{\nu/\mu}}=\int^\infty_0\exp(-zx^\mu)x^{\nu-1}\dd x,$$
with $\Re \nu>0, \mu>0$, and $\Re z>0$, one can further simplify the above equation \eqref{Pe1} as
\begin{eqnarray}
\begin{aligned}
    P_\text{exc,final}
    &=\frac{g_0^2}{\alpha^2}\abs{\bigg(\frac{1}{\alpha T_d}-\frac{i\omega_p}{\alpha}\bigg)^{-1-i\omega_q/\alpha}\Gamma\bigg[1+i\frac{\omega_q}{\alpha}\bigg]}^2
    \\
    &=\frac{g_0^2}{\alpha^2}\abs{(Ae^{-i\theta})^{-1-i\omega_q/\alpha}\bigg(i\frac{\omega_q}{\alpha}\bigg)\Gamma\bigg[i\frac{\omega_q}{\alpha}\bigg]}^2
    \\
    &=\frac{g_0^2}{\alpha^2}A^{-2}e^{-2\theta\omega_q/\alpha}\bigg(\frac{\omega_q}{\alpha}\bigg)^2\frac{\pi}{\frac{\omega_q}{\alpha}\sinh\big(\frac{\pi\omega_q}{\alpha}\big)}
    \\
    &=\frac{g_0^2}{\omega_p^2+T_d^{-2}} \frac{2\pi\omega_q}{\alpha} \frac{\exp[(\pi-2\theta)\frac{\omega_q}{\alpha}]}{\exp({\frac{2\pi\omega_q}{\alpha}})-1},
\end{aligned}
\label{Eq:P_excfinal}
\end{eqnarray}
where $A=\frac{1}{\alpha}\sqrt{\omega_p^2+T_d^{-2}}$, and $\tan\theta=\omega_p T_d$ with $\theta\in[0, \frac{\pi}{2}]$.
Similarly, one can also find the final probability of emission from $\ket{e,0}$ to $\ket{g,1}$, which is given by
\begin{eqnarray}
\begin{aligned}
    P_\text{emi,final}
    &=\abs{\int_0^\infty\,\dd t \bra{g,1}\hat{V}^\text{M}(t)\ket{e,0}}^2
    \\
    &=g_0^2\abs{\int^\infty_{0}\dd t e^{-t/T_d}e^{i\omega_pt}e^{-i\omega_q\ln(\alpha t)/\alpha}}^2
    \\
    &=g_0^2\abs{\int^\infty_{0}\dd t e^{-(1/T_d-i\omega_p)t}(\alpha t)^{-i\omega_q/\alpha}}^2
    \\
    &=\frac{g_0^2}{\omega_p^2+T_d^{-2}} \frac{2\pi\omega_q}{\alpha} \frac{\exp[(2\theta-\pi)\frac{\omega_q}{\alpha}]}{1-\exp(-{\frac{2\pi\omega_q}{\alpha}})}.
\end{aligned}
\label{Eq:P_emifinal}
\end{eqnarray}
Furthermore, if we assume the damping time $T_d\to\infty$, typically $\omega_p T_d\gg1$, then we can find $\theta\to\pi/2$ and Eqs.~\eqref{Eq:P_excfinal} and \eqref{Eq:P_emifinal} could be approximated as Eq.~(4) in the main text, namely
\begin{equation}
\begin{aligned}
    P_\text{exc,final}
    &= \frac{g_0^2}{\omega_p^2}\frac{2\pi\omega_q}{\alpha}
        \times\frac{1}{\exp(\frac{2\pi\omega_q}{\alpha})-1},\\
    P_\text{emi,final}
    &= \frac{g_0^2}{\omega_p^2}\frac{2\pi\omega_q}{\alpha}
        \times\frac{1}{1-\exp(-\frac{2\pi\omega_q}{\alpha})}.
    \label{Eq:P_final}
\end{aligned}
\end{equation}
The Planck factor in Eq.~\eqref{Eq:P_final} indicates the thermalization property~\cite{ben-benjamin_unruh_2019} in the time-like Unruh effect.

In the experiment we note that, in such case, the Hamiltonian in Eq.~\eqref{Eq:H_M_theory} turns to be infinity when $t \to 0$, which is challenging in experimental realization.
To solve this issue, we determine the beginning time of evolution as $\alpha t=C$, which corresponds to the time $\tau=0$ in the FP coordinate.
For convenience, we slightly modify the spin term in Eq.~\eqref{Eq:H_M_theory} through a time translation, i.e., $t\rightarrow t+C/\alpha$.
In such case, we have $\alpha t+C=e^{\alpha\tau}$ and the Hamiltonian in Eq.~\eqref{Eq:H_M_theory} is revised to be
\begin{equation}
    \label{Eq:H_M_exp}
    \hat{H}^\text{M}_\text{exp}
    = \frac{1}{\alpha t + C} \frac{\omega_q}{2} \hat{\sigma}_z
            + \omega_p \hat{b}^\dagger\hat{b}
            + g_0 \chi(t) \hat{\sigma}_x (\hat{b} + \hat{b}^\dagger).
\end{equation}
Note that, Eq.~\eqref{Eq:H_M_exp} is universal -- specially when $C=0$ it returns to be Eq.~\eqref{Eq:H_M_theory}.
Here, the spin energy spacing is $\omega_q/C$ originally ($t=0$) and continuously scaled by $1/(\alpha t+C)$.

Then we discuss the corresponding results caused by the time translation.
First, note that Eqs.~\eqref{Eq:P_excfinal} and \eqref{Eq:P_emifinal} are calculated based on $C=0$.
Considering the general case, i.e., replacing $\alpha t$ with $\alpha t + C$ and keeping $\chi(t)=\exp(-t/T_d)$, Eq.~\eqref{Eq:P_excfinal} should be revised as
\begin{equation}
\begin{aligned}
    P_\text{exc,final}
    &=g_0^2\abs{\int^\infty_{0}\dd t e^{-(1/T_d-i\omega_p)t}(\alpha t+C)^{i\omega_q/\alpha}}^2
    \\
    &=g_0^2 \exp({\frac{2C}{\alpha T_d}}) \\
    &\qquad \times \abs{\int^\infty_{C/\alpha}\dd t e^{-(1/T_d-i\omega_p)t}(\alpha t)^{i\omega_q/\alpha}}^2
    .
\end{aligned}
\end{equation}
From the definition of upper incomplete Gamma function,$$\frac{1}{\mu}\Gamma{\bigg(}\frac{\nu}{\mu},s{\bigg)} \frac{1}{z^{\nu/\mu}} = \int^\infty_s\exp(-zx^\mu)x^{\nu-1}\dd x,$$
with $\Re \nu>0, \mu>0, s\ge0$, and $\Re z>0$, one has
\begin{equation}
\begin{aligned}
    P_\text{exc,final}
    &=\frac{g_0^2}{\frac{1}{T_d^2}+\omega_p^2} \bigg(\frac{\omega_q}{\alpha}\bigg)^2 \exp({\frac{2C}{\alpha T_d}-2\theta\frac{\omega_q}{\alpha}})
    \\
    &\qquad \times \abs{\Gamma\bigg[1+i\frac{\omega_q}{\alpha},C\bigg]}^2
    .
\end{aligned}
\label{Eq:P_excfinal_revised}
\end{equation}
Obviously, when $C=0$ it returns to be Eq.~\eqref{Eq:P_excfinal}, and one can also obtain the corresponding result similarly for the emission process, that is
\begin{equation}
\begin{aligned}
    P_\text{emi,final}
    &=\frac{g_0^2}{\frac{1}{T_d^2}+\omega_p^2} \bigg(\frac{\omega_q}{\alpha}\bigg)^2 \exp({\frac{2C}{\alpha T_d}}+2\theta\frac{\omega_q}{\alpha})
    \\
    &\qquad \times \abs{\Gamma\bigg[1-i\frac{\omega_q}{\alpha},C\bigg]}^2
    .
\end{aligned}
\label{Eq:P_emifinal_revised}
\end{equation}
However, the incomplete Gamma function could not be analytically calculated.
Thus we will calculate it numerically after determining exact parameters (see following part).
But we have a qualitative conclusion, that is, the smaller the value of C (approaching zero) and the greater the acceleration, the closer the two results converge.

Actually in the experiment, we ignore the phonon state and merely detect the spin populations $P_g/P_e$, which can be expressed as
\begin{equation}
\begin{aligned}
    P_{g/e}
    &=\sum_n{\abs{\int\,\dd t \bra{g/e,n}\hat{V}^\text{M}(t)\ket{\psi_i}}^2}
    .
\end{aligned}
\end{equation}
For excitation process, the initial state is $\ket{\psi_i}=\ket{g,0}$ and $\bra{e,n}\hat{V}^\text{M}(t)\ket{\psi_i}\neq 0$ only when $n=1$.
Thus we get
\begin{equation}
\label{Eq:P_e}
\begin{aligned}
    P_{e}
    &=\sum_n{\abs{\int\,\dd t \bra{e,n}\hat{V}^\text{M}(t)\ket{g,0}}^2}
    \\
    &=\abs{\int\,\dd t \bra{e,1}\hat{V}^\text{M}(t)\ket{g,0}}^2
    = P_\text{exc}
    ,
\end{aligned}
\end{equation}
equal to Eq.~\eqref{Eq:P_excfinal}.
Similarly for emission process, the initial state is $\ket{\psi_i}=\ket{g,0}$ and then we get
\begin{equation}
\label{Eq:P_g}
\begin{aligned}
    P_{g}
    &=\sum_n{\abs{\int\,\dd t \bra{g,n}\hat{V}^\text{M}(t)\ket{e,0}}^2}
    \\
    &=\abs{\int\,\dd t \bra{g,1}\hat{V}^\text{M}(t)\ket{e,0}}^2
    = P_\text{emi}
    ,
\end{aligned}
\end{equation}
equal to Eq.~\eqref{Eq:P_emifinal}.
However, experimental imperfections lead to some acceptable deviations, which is detailed in the corresponding section below.

\section{Hamiltonian construction}
Considering a universal model of a trapped ion, the total Hamiltonian reads $\hat{H}_\text{total} = \hat{H}_\text{spin} + \hat{H}_\text{phonon} + \hat{H}_\text{laser}$.
Here, $\hat{H}_\text{spin}$ is the inertial Hamiltonian of the ion, and we only consider two inner levels, corresponding to $\hat{H}_\text{spin} = \omega_0 \hat{\sigma}_z / 2$.
$\hat{H}_\text{phonon}$ is the motional Hamiltonian of the ion, and we only consider the axial ($z$ direction) motion, corresponding to $\hat{H}_\text{phonon} = \omega_z \hat{b}^\dagger \hat{b}$.
$\hat{H}_\text{laser}$, representing the interaction with laser(s), usually reads $\sum_j{\Omega_j(\exp[i(\omega_j t-k_j \hat{z} + \phi_j)] + \text{H.c.})}$, where $\Omega_j,\omega_j,k_j,\phi_j$ represent the intensity, frequency, wave vector and phase of the $j$-th laser, respectively, and $i$ is the imaginary unit.
The position operator $\hat{z}$ could be expressed as $\hat{z}=z_0(\hat{b}+\hat{b}^\dagger)$.
Thus, the total Hamiltonian in our model reads
\begin{equation}
\begin{aligned}
    \hat{H}_\text{total}
    &= \frac{\omega_0}{2} \hat{\sigma}_z + \omega_z \hat{b}^\dagger \hat{b}\\
    &+ \sum_{j}{\Omega_j \bigg(\exp\bigg[i\big(\omega_j t-k_j z_0(\hat{b}+\hat{b}^\dagger) + \phi_j\big)\bigg] + \text{H.c.}\bigg) }.
\end{aligned}
\label{Eq:H_ion}
\end{equation}

To construct the Hamiltonian in Eq.~(6) in the main text using trapped ion, we apply two customized lasers, with frequencies
\begin{equation}
\label{Eq:laser frequency}
\omega_\text{1,2}=\bigg[\omega_0-\omega_q\frac{\ln{(\alpha t+C)}}{\alpha t}\bigg]\pm(\omega_z-\omega_p),
\end{equation}
the same time-dependent intensity $\Omega_1=\Omega_2=\Omega_0\chi(t)$ and phase $\phi_1=\phi_2=0$.
Then we rotate the Hamiltonian by
$$\hat{H}_\text{int} = i\frac{\dd \hat{U}_0^\dagger}{\dd t}\hat{U}_0+\hat{U}_0^\dagger \hat{H}_\text{total} \hat{U}_0$$
with the operator $\hat{U}_0=\exp(-i\int_0^{t}\hat{H}_0 \dd t)$, where
$$\hat{H}_0 = (\omega_z-\omega_p){\hat{b}^\dagger}\hat{b} - \frac{1}{2}(\omega_0 - \frac{\omega_q}{\alpha t+C})\hat{\sigma}_z.$$
Using rotating-wave approximation (RWA), we can finally get the constructed Hamiltonian as
\begin{equation}
\begin{aligned}
    \hat{H}_\text{int}
    &\approx \frac{1}{\alpha t+C}\frac{1}{2}\omega_q\hat{\sigma}_z + \omega_p\hat{b}^\dagger \hat{b}\\
    &\qquad + i\frac{\eta_1\Omega_1}{2} (\hat{\sigma}_+{\hat{b}^\dagger} + \hat{\sigma}_-\hat{b})
     + i\frac{\eta_2\Omega_2}{2}(\hat{\sigma}_+\hat{b}+\hat{\sigma}_-{\hat{b}^\dagger})\\
    &\approx \frac{1}{\alpha t+C}\frac{1}{2}\omega_q\hat{\sigma}_z + \omega_p{\hat{b}^\dagger}\hat{b}+g_0\chi(t)\hat{\sigma}_x ({\hat{b}+\hat{b}^\dagger}).
\end{aligned}
\end{equation}
Here, $\eta_j=k_j z_0 \approx \eta$ is the Lamb-Dicke parameter and the coupling strength $g_0=\eta\Omega_0/2$ (ignoring the imaginary term).
Thus we construct the Hamiltonian in Eq.~(6) in the main text using a single trapped ion with custom laser.

\section{Experimental parameter settings}
The experimental realization of Eq.~\eqref{Eq:H_M_exp} requires carefully constrained parameters that simultaneously satisfy:
(i) theoretical validity conditions for the detector-field model,
(ii) implementability within trapped-ion hardware limitations, and
(iii) measurability given decoherence timescales.
We systematically address these constraints below, with key parameters summarized in Table~S1.

\begin{table*}[htbp]
\centering
\caption{Key experimental parameters for timelike Unruh effect demonstration}
\label{tab:experimental_parameters}
\scriptsize 
\begin{tblr}{
  width = 0.8\linewidth,
  colspec = {
    l
    X[1,c]  
    X[1,c]  
    Q[0.3\linewidth, l]  
  },
  row{1} = {font=\bfseries}, 
  row{even} = {belowsep=0.5ex}, 
  row{odd} = {belowsep=0.5ex}, 
  hline{1} = {1.5pt}, 
  hline{2} = {0.5pt}, 
  hline{Z} = {1.5pt}, 
  cells = {font=\scriptsize} 
}
Parameter & Symbol & Value & Constraint Basis \\
Spin freq. & $\omega_q$ & $2\pi \times 200\ \text{kHz}$ & 
$\begin{aligned} 
  &\bullet \text{Observability } \omega_q \ge 4\omega_p \\ 
  &\bullet \text{RWA validity } \omega_q \ll \omega_z 
\end{aligned}$ \\
Phonon freq. & $\omega_p$ & 
$\begin{aligned}
  & 2\pi \times 25\ \text{kHz} \\ 
  & 2\pi \times 50\ \text{kHz}
\end{aligned}$
& 
$\begin{aligned}
  & \bullet \text{Measurability } P_\text{final}>0.02\\ 
  & \bullet \text{Approx. condition } P_\text{final}<0.1\ll1
\end{aligned}$
\\
Effective acceleration & $\alpha$ &  $10^3-10^6\ \text{kHz}$ &
$\begin{aligned}
  & \bullet \beta=\alpha/(2\pi\omega_q)\in[10^{-1},10^2]
\end{aligned}$
\\
Coupling strength & $g_0$ &  $2\pi \times 5\ \text{kHz}$ &
$\begin{aligned}
  & \bullet \text{Laser power}
\end{aligned}$
\\
Damping time & $T_d$ & $0.2\ \text{ms}$ & 
$\begin{aligned} 
  & \bullet \omega_p T_d = 5\pi \gg 1\\ 
  & \bullet T_d \ll t_\text{max}
\end{aligned}$
\\
Max. evo. time & $t_{\text{max}}$ & $1\ \text{ms}$ & 
$\begin{aligned}
  & \bullet \text{Coherence time \& heating rate}
\end{aligned}$
\\
Temp. tran. const. & $C$ & $1$ & 
$\begin{aligned}
  & \bullet \text{RWA validity}
\end{aligned}$
\\
\end{tblr}
\end{table*}

\textit{RWA Validity Constraints}---%
To prevent laser frequencies (Eq.~\eqref{Eq:laser frequency}) from approaching spin resonance ($\omega_{1,2}(t) \approx \omega_0$)—which would violate the RWA—the temporal translation constant $C$ cannot be arbitrarily small.
We therefore fix $C=1$ and constrain $\omega_q, \omega_p \ll \omega_z = 2\pi \times 1.096~\text{MHz}$.

\textit{Observability Conditions}---%
While the ideal Hamiltonian (Eq.~\eqref{Eq:H_M_theory}) yields $\omega'_q = \omega_q/(\alpha t) \in [0, \infty)$, the experimental implementation (Eq.~\eqref{Eq:H_M_exp}) bounds $\omega'_q \in [0, \omega_q]$.
Numerical simulations establish equivalence when $\omega_q > k\omega_p$ ($k \sim \mathcal{O}(1)$).
Furthermore, note that the emission process constitutes a Landau-Zener transition~\cite{marco_simulating_2012} requiring resonant coupling ($\omega'_q = \omega_p$) at critical times.

\textit{First-Order Approximation Validity}---%
The condition $P_{fi} \ll 1$ (Eq.~\eqref{Eq:P_fi}) necessitates weak coupling $g_0/\omega_p \ll 1$, as $P_\text{final} \propto (g_0/\omega_p)^2$ (Eq.~\eqref{Eq:P_final}).
Given laser power limitations, we set $g_0 = 2\pi \times 5~\text{kHz}$.
This yields $\omega_q/(2\pi) = 200~\text{kHz}$ and $\omega_p/(2\pi) = 25~\text{kHz}$, satisfying $g_0/\omega_p = 0.2,\omega_q/\omega_p = 8$ and meeting all prior constraints.

\begin{figure}[htbp]
	\begin{center}
	\includegraphics[width=1\columnwidth]{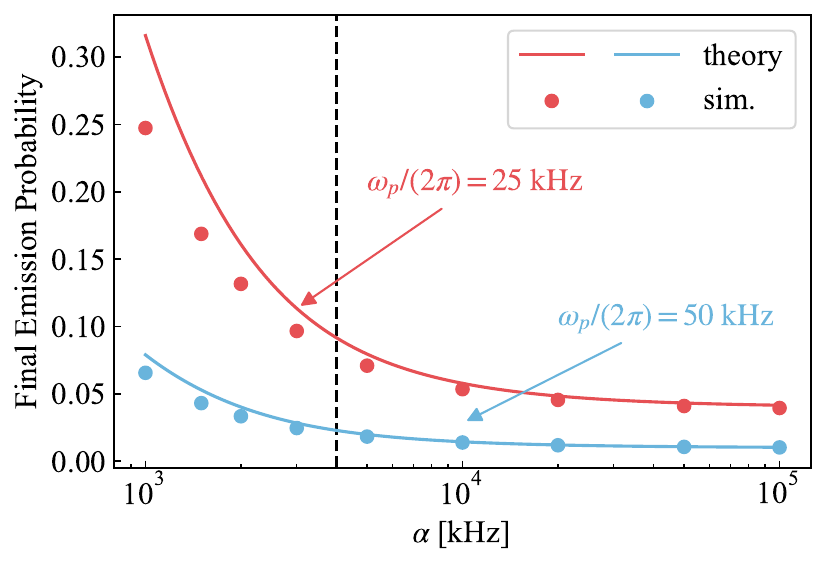}
	\caption{
        Comparison between two simulation results with parameter $\omega_p/(2\pi)=25~\text{kHz}$ (red) and 50~kHz (blue).
        Other parameters are shown in Table~\ref{tab:experimental_parameters}.
        The solid lines denote theoretical prediction~\eqref{Eq:P_final} while the dots represent simulation results.
        The vertical dashed line corresponds to zone I/II division in Fig.~3 (main text).
        }
	\label{fig:Comparison of omega_p}
	\end{center}
\end{figure}

For emission dynamics under low acceleration, the first-order approximation fails at $\omega_p = 25~\text{kHz}$, causing significant theoretical deviations (Fig.~\ref{fig:Comparison of omega_p}).
Although increasing to $\omega_p = 50~\text{kHz}$ improves agreement, it introduces experimental challenges due to heightened acceleration requirements.
We therefore implement: $\omega_p = 50~\text{kHz}$ for low-acceleration regimes ($\alpha < \alpha_c$) and $\omega_p = 25~\text{kHz}$ for high-acceleration regimes ($\alpha > \alpha_c$), where $\alpha_c$ (dashed boundary, Fig.~\ref{fig:Comparison of omega_p}) corresponds to the zone I/II division in Fig.~3 (main text).
This maintains $P_\text{emi,final} < 0.1$—preserving approximation validity—without affecting the effective probability $P_\text{eff} = (\omega_p^2/g_0^2) P_\text{final}$.

\textit{Damping and evolution time}---%
As discussed above, the approximation $P_{\text{final}}$ will reduce to Eq.~\eqref{Eq:P_final} when $\omega_p T_d \gg 1$. Moreover, experimental heating rates and coherence times constrain the maximal evolution time $t_\text{max}$, and it should be several times greater than the damping time $T_d$ to observe the steady state.
These lead to our experimental parameter choice: $T_d=0.2~\text{ms}, t_\text{max}\simeq5T_d = 1~\text{ms}$, which yields $\omega_p T_d = 5\pi \gg 1$, preserving the corresponding validity.

\textit{Parameter validation}---%
To verify the validation of the parameter selection mentioned above, we compare the revised results (Eqs.~\eqref{Eq:P_excfinal_revised}\eqref{Eq:P_emifinal_revised}) and the original results (Eq.~\eqref{Eq:P_final}).
Moreover, we set two values of $C$ to analyze the influence of the time translation.
Results with $C=0$ (crosses) match theoretical predictions well while those with $C=1$ (stars) show minor deviations, negligible versus experimental noise (see Fig.~3 in the main text).
This confirms our parameter choice preserves physical consistency.
Note that, the $C=0$ results could not be simulated due to the infinite spin frequency at $t=0$ but could be calculated through symbolic computation.

\begin{figure}[htbp]
	\begin{center}
	\includegraphics[width=1\columnwidth]{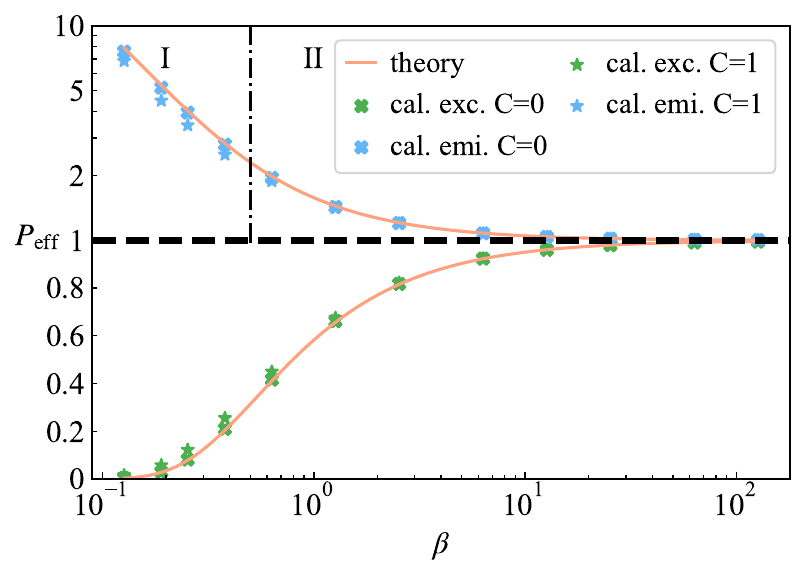}
	\caption{
        Comparison of theoretical predictions and calculated results with $C=0$ and $C=1$.
        The parameter settings of points are the same as those in Fig.~2 (main text).
        }
	\label{fig:time translation P}
	\end{center}
\end{figure}

\section{Numerical simulation and experimental imperfections}

\begin{figure}[t!]
	\begin{center}
	\includegraphics[width=\columnwidth]{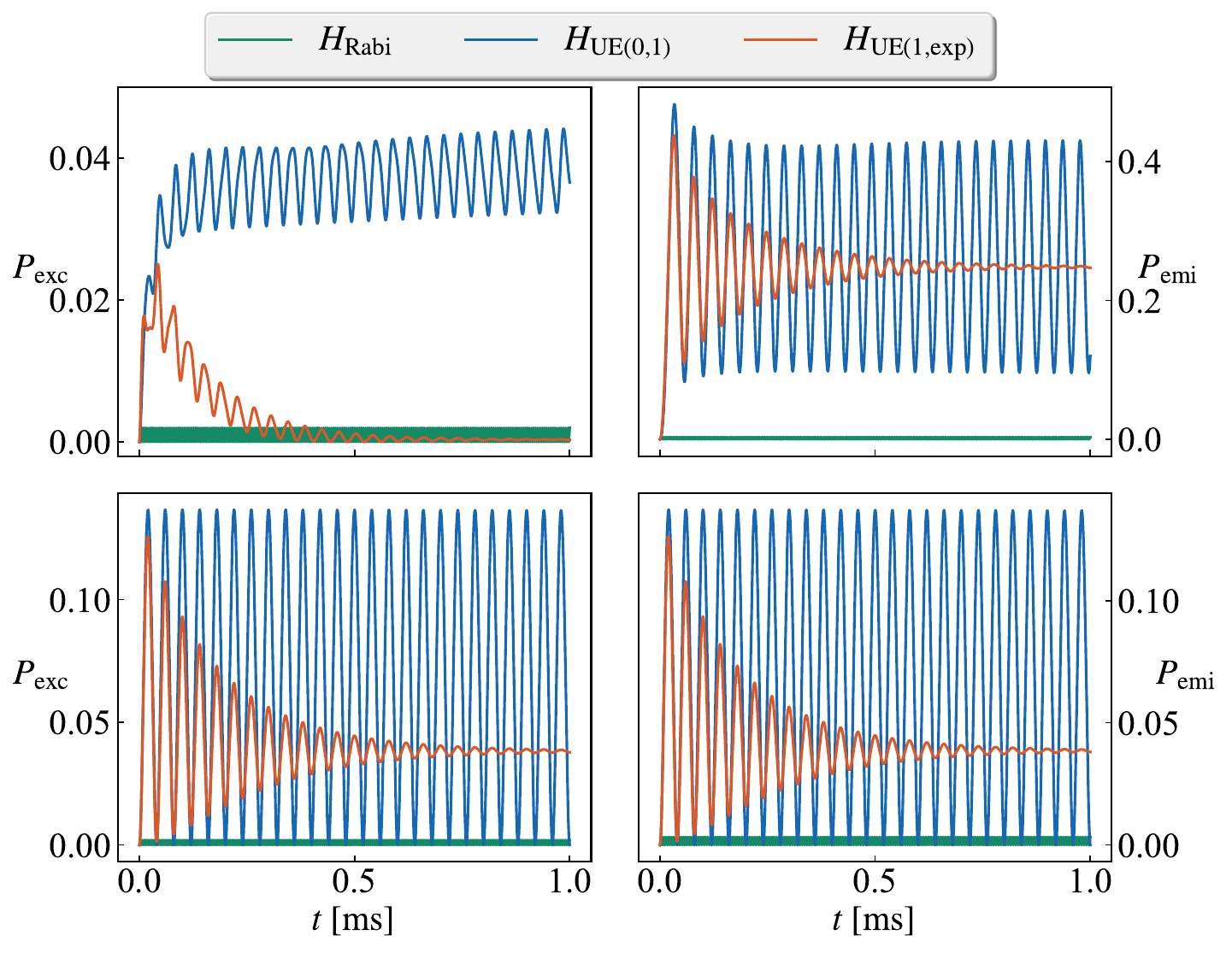}
	\caption{
        Comparison of $H_\text{Rabi}$, $H_\text{UE(0,1)}$ and $H_\text{UE(1,exp)}$ model.
        We consider $2\times2$ conditions: excitation (left) or emission (right) process and low (top) or high (bottom) acceleration as $\alpha=1~\text{MHz}$ or $1~\text{GHz}$.
        Other parameters for all the models are setting as $\omega_q/2\pi = 200~\text{kHz},\quad \omega_p/2\pi = 25~\text{kHz},\quad g_0/2\pi = 5~\text{kHz}$.
        The total evolution time is $t=5T_d=1~\text{ms}$.
        }
	\label{fig:Comparison}
	\end{center}
\end{figure}

In order to better understand the system dynamics dominated by Eq.~\eqref{Eq:H_M_exp}, we study it with various cases in the following.
Specifically, we will consider three models 
with corresponding Hamiltonians respectively given by 
\begin{align}
\hat{H}_\text{Rabi} &= \frac{\omega_q}{2}\hat{\sigma}_z + \omega_p\hat{b}^\dagger \hat{b} + g_0 \hat{\sigma}_x ({\hat{b}+\hat{b}^\dagger}),\label{Eq:H_Rabi}
\\
\hat{H}_\text{UE(0,1)} &= \frac{1}{\alpha t}\frac{\omega_q}{2}\hat{\sigma}_z + \omega_p\hat{b}^\dagger \hat{b} + g_0 \hat{\sigma}_x ({\hat{b}+\hat{b}^\dagger}),\label{Eq:H_UE}
\\
\hat{H}_\text{UE(1,exp)} &= \frac{1}{\alpha t+1}\frac{\omega_q}{2}\hat{\sigma}_z + \omega_p\hat{b}^\dagger \hat{b} + g_0 e^{-t/T_d} \hat{\sigma}_x ({\hat{b}+\hat{b}^\dagger}).\label{Eq:H_UE_exp}
\end{align}
Case 1, denoted by $H_\text{Rabi}$, is actually a Rabi model and it corresponds to the condition of $\alpha=0, C=1, \chi(t)\equiv1$ in Eq.~\eqref{Eq:H_M_exp}.
Case 2, denoted by $H_\text{UE(0,1)}$, corresponds to the condition of $C=0, \chi(t)\equiv1$ in Eq.~\eqref{Eq:H_M_exp} and it represents that an accelerated Unruh-DeWitt detector model with a constant coupling in the future light cone in the laboratory framework.
Case 3, denoted by $H_\text{UE(1,exp)}$, is exactly the Hamiltonian we construct in the experiment with parameters $C=1, \chi(t)=\exp(-t/T_d)$.
We take various parameters and show the relevant excitation and emission dynamics in Fig.~\ref{fig:Comparison}.

Comparing Hamiltonian in
Eq.~\eqref{Eq:H_Rabi} with that in Eq.~\eqref{Eq:H_UE}, the only difference is the scaling term $1/\alpha t$.
Due to the non-resonance in the Rabi model ($\omega_p\neq\omega_p$), the evolution is trivial and no transition happens.
However, the scaling term in Eq.~\eqref{Eq:H_UE}, due to the Doppler shift caused by acceleration, leads to a non-trivial dynamics, and in the long-time limit, the system settles into a steady oscillatory state.
In the long-time limit, it is actually a detuned Rabi oscillation and the corresponding Hamiltonian is $\omega_p\hat{b}^\dagger \hat{b} + g_0 \hat{\sigma}_x (\hat{b}+\hat{b}^\dagger)$.
For the $H_\text{UE(1,exp)}$ model in Eq.~\eqref{Eq:H_UE_exp}, we find that its corresponding evolution in the long-time limit is steady without any oscillation.
This results from the switching function of exponential damping, and in such case only $\omega_p\hat{b}^\dagger \hat{b}$ term in the Hamiltonian is left in the end.
Therefore, in such case we can think in the end the detector and the field are decoupled such that the final populations of the detector and number of the photon are the thermal equilibrium ones purely resulting from the Unruh effect.

For Fig.~2 (main text), we simulate the models under conditions with and without noise.
For the ideal condition (denoted by ‘sim. ideal’), the simulated Hamiltonian is just Eq.~\eqref{Eq:H_M_exp} and without any noise.
After observing the discrepancy between experimental results and ideal simulation, we suspect that it is due to the imperfection of the experiment.
We also simulate the real Hamiltonian Eq.~\eqref{Eq:H_ion} by considering possible experimental noise (denoted by ‘sim. noise’).

First, without noise the simulation results of Eq.~\eqref{Eq:H_ion} are consistent with those of Eq.~\eqref{Eq:H_M_exp}, which indicates that the rotation-wave approximation holds.
Considering actual situation, we introduce different experimental imperfections, and finally find that the key factors are preparation of initial phonon state and relative phase between two lasers.
Before the experiment, imperfect phonon cooling will lead to that the initial phonon state is not the ground state $\ket{n=0}$, and we expect but a thermal state $\hat{\rho}_{\Bar{n}}$ instead, where $\Bar{n}$ is the average phonon number after cooling.
More explicitly, the real initial state is $\ket{g}\bra{g}\otimes\hat{\rho}_{\Bar{n}}$ or $\ket{e}\bra{e}\otimes\hat{\rho}_{\Bar{n}}$ and meanwhile the unsteady cooling efficiency leads to that $\Bar{n}$ fluctuates approximately between 0.02 and 0.07 (experimentally measured) during experimental repetitions.
Due to the small $\Bar{n}$, the experimental results deviate from theoretical predictions slightly according to Eqs.~\eqref{Eq:P_e} and \eqref{Eq:P_g}.
On the other hand, although we expect to fix the relative phase of two lasers, in the actual experiment it is hard to control due to the unsteady switching time of ascoutic-optic modulator.
Moreover, we consider spin and phonon decoherence and heating rate, which has few contributions to residual though.
These noise components constitute the error bands illustrated in Fig.~2.

Similarly in Fig.~3 (main text), two error bands are obtained by simulating Eq.~\eqref{Eq:H_ion} with various noises and extracting the bound of final state.

All numerical simulations mentioned in the main text and supplementary materials are completed using \textit{qutip} library in python~\cite{johansson-qutip-2012}.



\begin{thebibliography}{100}

\bibitem{unruh_notes_1976}
W. G. Unruh, Notes on black-hole evaporation,~\href{https://journals.aps.org/prd/abstract/10.1103/PhysRevD.14.870}
{Phys. Rev. D {\bf 14}, 870 (1976)}.

\bibitem{crispino_unruh_2008}
Lu\'is C. B. Crispino, Atsushi Higuchi, and George E. A. Matsas, The Unruh effect and its applications,~\href{https://journals.aps.org/rmp/abstract/10.1103/RevModPhys.80.787}
{Rev. Mod. Phys. {\bf 80}, 787 (2008)}.

\bibitem{birrell_quantum_1982}
N. D. Birrell and P. Davies, \emph{Quantum Fields in Curved Space}, {(Cambridge University Press, Cambridge, England, 1984)}.

\bibitem{bell_electrons_1983}
J. S. Bell and J. M. Leinaas, Electrons as accelerated thermometers,~\href{https://doi.org/10.1016/0550-3213(83)90601-6}
{Nuclear Physics B, {\bf 212}, 131-150 (1983)}.

\bibitem{yablonovitch_accelerating_1989}
E. Yablonovitch, Accelerating reference frame for electromagnetic waves in a rapidly growing plasma: Unruh-Davies-Fulling-DeWitt radiation and the nonadiabatic Casimir effect,~\href{https://doi.org/10.1103/PhysRevLett.62.1742}
{Phys. Rev. Lett. {\bf 62}, 1742 (1989)}.

\bibitem{chen_testing_1999}
Pisin Chen and Toshi Tajima, Testing Unruh radiation with ultraintense lasers,~\href{https://doi.org/10.1103/PhysRevLett.83.256}
{Phys. Rev. Lett. {\bf 83}, 256 (1999)}.

\bibitem{schutzhold_signatures_2006}
Ralf Sch\"utzhold, Gernot Schaller, and Dietrich Habs, Signatures of the Unruh effect from electrons accelerated by ultrastrong laser fields,~\href{https://doi.org/10.1103/PhysRevLett.97.121302}
{Phys. Rev. Lett. {\bf 97}, 121302 (2006)}.

\bibitem{schutzhold_tabletop_2008}
Ralf Sch\"utzhold, Gernot Schaller, and Dietrich Habs, Tabletop creation of entangled multi-keV photon pairs and the Unruh effect,~\href{https://doi.org/10.1103/PhysRevLett.100.091301}
{Phys. Rev. Lett. {\bf 100}, 091301 (2008)}.

\bibitem{martin-martinez_using_2011}
Eduardo Mart\'in-Mart\'inez, Ivette Fuentes, and Robert B. Mann, Using Berry’s Phase to detect the Unruh effect at lower accelerations,~\href{https://doi.org/10.1103/PhysRevLett.107.131301}
{Phys. Rev. Lett. {\bf 107}, 131301 (2011)}.

\bibitem{guedes_spectra_2019}
T. L. M. Guedes, M. Kizmann, D. V. Seletskiy, A. Leitenstorfer, Guido Burkard, and A. S. Moskalenko, Spectra of ultrabroadband squeezed pulses and the finite-time Unruh-Davies effect,~\href{https://doi.org/10.1103/PhysRevLett.122.053604}
{Phys. Rev. Lett. {\bf 122}, 053604 (2019)}.

\bibitem{lochan_detecting_2020}
Kinjalk Lochan, Hendrik Ulbricht, Andrea Vinante, and Sandeep K. Goyal, Detecting acceleration-enhanced vacuum fluctuations with atoms inside a cavity,~\href{https://doi.org/10.1103/PhysRevLett.125.241301}
{Phys. Rev. Lett. {\bf 125}, 241301 (2020)}.

\bibitem{gooding_interferometric_2020}
Cisco Gooding, Steffen Biermann, Sebastian Erne, Jorma Louko, William G. Unruh, J\"org Schmiedmayer, and Silke Weinfurtner, Interferometric Unruh detectors for Bose-Einstein condensates,~\href{https://doi.org/10.1103/PhysRevLett.125.213603}
{Phys. Rev. Lett. {\bf 125}, 213603 (2020)}.

\bibitem{arya_geometric_2022}
Navdeep Arya, Vikash Mittal, Kinjalk Lochan, and Sandeep K. Goyal, Geometric phase assisted observation of noninertial cavity-QED effects,~\href{https://doi.org/10.1103/PhysRevD.106.045011}
{Phys. Rev. D {\bf 106}, 045011 (2022)}.

\bibitem{tian_probing_2022}
Zehua Tian, Longhao Wu, Liang Zhang, Jiliang Jing, and Jiangfeng Du, Probing Lorentz-invariance-violation-induced nonthermal Unruh effect in quasi-two-dimensional dipolar condensates,~\href{https://doi.org/10.1103/PhysRevD.106.L061701}
{Phys. Rev. D {\bf 106}, L061701 (2022)}.

\bibitem{stargen_cavity_2022}
D. Jaffino Stargen and Kinjalk Lochan, Cavity optimization for Unruh effect at small accelerations,~\href{https://doi.org/10.1103/PhysRevLett.129.111303}
{Phys. Rev. Lett. {\bf 129}, 111303 (2022)}.

\bibitem{arya_lamb_2023}
Navdeep Arya and Sandeep K. Goyal, Lamb shift as a witness for quantum noninertial effects,~\href{https://doi.org/10.1103/PhysRevD.108.085011}
{Phys. Rev. D {\bf 108}, 085011 (2023)}.

\bibitem{BEC}
H. Jiazhong, L. Feng, Z. Zhang, C. Chin, Quantum simulation of Unruh radiation,~\href{https://doi.org/10.1038/s41567-019-0537-1}
{Nat. Phys. {\bf 15}, 785–789 (2019)}.

\bibitem{NMR}
F.Z. Jin, H.W. Chen, X. Rong, H. Zhou, M.J. Shi, Q. Zhang, J. ChenYong, Y.F. Cai, S.L. Luo, X.H. Peng, J.F. Du, Experimental simulation of the Unruh effect on an NMR quantum simulator,~\href{https://doi.org/10.1007/s11433-016-5779-7}
{Sci. China Phys. Mech. Astron. {\bf 59}, 630302 (2016)}.

\bibitem{olson_entanglement_2011}
S. Jay Olson and Timothy C. Ralph, Entanglement between the future and the past in the quantum vacuum,~\href{https://doi.org/10.1103/PhysRevLett.106.110404}
{Phys. Rev. Lett. {\bf 106}, 110404 (2011)}.

\bibitem{olson_extraction_2012}
S. Jay Olson and Timothy C. Ralph, Extraction of timelike entanglement from the quantum vacuum,~\href{https://doi.org/10.1103/PhysRevA.85.012306}
{Phys. Rev. A {\bf 85}, 012306 (2012)}.

\bibitem{higuchi_entanglement_2017}
Atsushi Higuchi, Satoshi Iso, Kazushige Ueda, and Kazuhiro Yamamoto, Entanglement of the vacuum between left, right, future, and past: The origin of entanglement-induced quantum radiation,~\href{https://doi.org/10.1103/PhysRevD.96.083531}
{Phys. Rev. D {\bf 96}, 083531 (2017)}.

\bibitem{ueda_entanglement_2021}
Kazushige Ueda, Atsushi Higuchi, Kazuhiro Yamamoto, Ar Rohim, and Yue Nan, Entanglement of the vacuum between left, right, future, and past: Dirac spinor in Rindler and Kasner spaces,~\href{https://doi.org/10.1103/PhysRevD.103.125005}
{Phys. Rev. D {\bf 103}, 125005 (2021)}.

\bibitem{quach_berry_2022}
James Q. Quach, Timothy C. Ralph, and William J. Munro, Berry phase from the entanglement of future and past light cones: Detecting the timelike Unruh effect,~\href{https://doi.org/10.1103/PhysRevLett.129.160401}
{Phys. Rev. Lett. {\bf 129}, 160401 (2022)}.

\bibitem{tian_using_2023}
Zehua Tian and Jiliang Jing, Using nanokelvin quantum thermometry to detect timelike Unruh effect in a Bose–Einstein condensate,~\href{https://doi.org/10.1140/epjc/s10052-023-12191-6}
{Eur. Phys. J. C {\bf 83}, 1022 (2023)}.

\bibitem{leibfried_quantum_2003} D. Leibfried, R. Blatt, C. Monroe, and D. Wineland, Quantum dynamics of single trapped ions,~\href{https://doi.org/10.1103/RevModPhys.75.281}
{Rev. Mod. Phys. \textbf{75}, 281 (2003)}.

\bibitem{alsing-ion-2005}
Paul M. Alsing, Jonathan P. Dowling, and G. J. Milburn, Ion Trap Simulations of Quantum fields in an expanding universe,~\href{https://doi.org/10.1103/PhysRevLett.94.220401}
{Phys. Rev. Lett. \textbf{94}, 220401 (2005)}.

\bibitem{menicucci-single-2007}
Nicolas C. Menicucci and G. J. Milburn, Single trapped ion as a time-dependent harmonic oscillator,~\href{https://doi.org/10.1103/PhysRevA.76.052105}
{Phys. Rev. A \textbf{76}, 052105 (2007)}.

\bibitem{menicucci-simulating-2010}
N. C. Menicucci, S. Jay Olson, and G. J. Milburn, Simulating quantum effects of cosmological expansion using a static ion trap,~\href{https://doi.org/10.1088/1367-2630/12/9/095019}
{New J. Phys. \textbf{12}, 095019 (2010)}.

\bibitem{schutzhold-analogue-2007}
Ralf Sch\"utzhold, Michael Uhlmann, Lutz Petersen, Hector Schmitz, Axel Friedenauer, and Tobias Sch\"atz, Analogue of cosmological particle creation in an ion trap,~\href{https://doi.org/10.1103/PhysRevLett.99.201301}
{Phys. Rev. Lett. \textbf{99}, 201301 (2007)}.

\bibitem{wittemer-phonon-2019}
Matthias Wittemer, Frederick Hakelberg, Philip Kiefer, Jan-Philipp Schr\"oder, Christian Fey, Ralf Sch\"utzhold, Ulrich Warring, and Tobias Schaetz, Phonon pair creation by inflating quantum fluctuations in an ion trap,~\href{https://doi.org/10.1103/PhysRevLett.123.180502}
{Phys. Rev. Lett. \textbf{123}, 180502 (2019)}.

\bibitem{horstmann-hawking-2010}
B. Horstmann, B. Reznik, S. Fagnocchi, and J. I. Cirac, Hawking radiation from an acoustic black hole on an ion ring,~\href{https://doi.org/10.1103/PhysRevLett.104.250403}
{Phys. Rev. Lett. \textbf{104}, 250403 (2010)}.

\bibitem{horstmann-hawking-2011}
B. Horstmann, R. Schützhold, B. Reznik, S. Fagnocchi, and J. Ignacio Cirac, Hawking radiation on an ion ring in the quantum regime,~\href{https://doi.org/10.1088/1367-2630/13/4/045008}
{New J. Phys. \textbf{13}, 045008 (2011)}.

\bibitem{tian_testing_2022}
Zehua Tian, Yiheng Lin, Uwe R. Fischer and Jiangfeng Du, Testing the upper bound on the speed of scrambling with an analogue of Hawking radiation using trapped ions,~\href{https://doi.org/10.1140/epjc/s10052-022-10149-8}
{Eur. Phys. J. C {\bf 82}, 212 (2022)}.

\bibitem{junker-adiabatic-2002}
W. Junker and E. Schrohe, Adiabatic vacuum states on general spacetime manifolds: Definition, construction, and physical properties,~\href{https://doi.org/10.1007/s000230200001}
{Ann. Henri Poincaré \textbf{3}, 1113–1181 (2002)}.

\bibitem{louko-transition-2008}
Jorma Louko and Alejandro Satz, Transition rate of the Unruh–DeWitt detector in curved spacetime,~\href{https://doi.org/10.1088/0264-9381/25/5/055012}
{Class. Quant. Grav. \textbf{25}, 055012 (2008)}.

\bibitem{Alejandro-harvesting-2015}
Alejandro Pozas-Kerstjens and Eduardo Mart\'{\i}n-Mart\'{\i}nez, Harvesting correlations from the quantum vacuum,~\href{https://doi.org/10.1103/PhysRevD.92.064042}
{Phys. Rev. D \textbf{92}, 064042 (2015)}.

\bibitem{Eduardo-causality-2015}
Eduardo Mart\'{\i}n-Mart\'{\i}nez, Causality issues of particle detector models in QFT and quantum optics,~\href{https://doi.org/10.1103/PhysRevD.92.104019}
{Phys. Rev. D \textbf{92}, 104019 (2015)}.

\bibitem{BLHu-relativistic-2012}
B. L. Hu, Shih-Yuin Lin and Jorma Louko, Relativistic quantum information in detectors–field interactions,~\href{https://doi.org/10.1088/0264-9381/29/22/224005}
{Class. Quantum Grav. \textbf{29}, 224005 (2012)}.

\bibitem{Alejandro-Then-2007}
Alejandro Satz, Then again, how often does the Unruh–DeWitt detector click if we switch it carefully?,~\href{https://doi.org/10.1088/0264-9381/24/7/003}
{Class. Quantum Grav. \textbf{24}, 1719 (2007)}.

\bibitem{louko-how-2006}
Jorma Louko and Alejandro Satz, How often does the Unruh–DeWitt detector click? Regularization by a spatial profile,~\href{https://doi.org/10.1088/0264-9381/23/22/015}
{Class. Quantum Grav. \textbf{23}, 6321 (2006)}.

\bibitem{Soda_Acceleration_2022}
B. \ifmmode \check{S}\else \v{S}\fi{}oda, V. Sudhir, and A. Kempf, Acceleration-induced effects in stimulated light-matter interactions,~\href{https://doi.org/10.1103/PhysRevLett.128.163603}
{Phys. Rev. Lett. {\bf 128}, 163603 (2022)}.

\bibitem{SM} See Supplementary Material for details about theoretical induction, experimental parameter settings, numerical simulation and experimental imperfections.

\bibitem{ben-benjamin_unruh_2019}
J. S. Ben-Benjamin, M. O. Scully, S. A. Fulling, D. M. Lee, D. N. Page, A. A. Svidzinsky, M. S. Zubairy, M. J. Duff, R. Glauber, W. P. Schleich, and W. G. Unruh, Unruh acceleration radiation revisited,~\href{https://doi.org/10.1142/S0217751X19410057}
{Int. J. Mod. Phys. A \textbf{34}, 1941005 (2019)}.

\bibitem{pedernales_dirac_2018}
J. S. Pedernales, M. Beau, S. M. Pittman, I. L. Egusquiza, L. Lamata, E. Solano, and A. del Campo, Dirac equation in (1+1)-dimensional curved spacetime and the multiphoton quantum Rabi model,~\href{https://doi.org/10.1103/PhysRevLett.120.160403}
{Phys. Rev. Lett. {\bf 120}, 160403 (2018)}.



\bibitem{marco_simulating_2012}
Marco del Rey, Diego Porras, and Eduardo Mart\'{\i}n-Mart\'{\i}nez, Simulating accelerated atoms coupled to a quantum field,~\href{https://doi.org/10.1103/PhysRevA.85.022511}{Phys. Rev. A \textbf{85}, 022511 (2012)}.

\bibitem{johansson-qutip-2012}
J. R. Johansson, P. D. Nation, and Franco Nori, QuTiP: An open-source Python framework for the dynamics of open quantum systems,~\href{https://doi.org/10.1016/j.cpc.2012.02.021}{Comp. Phys. Comm. \textbf{183}, 1760–1772 (2012)}.

\end{thebibliography}
\end{document}